\documentclass[12pt]{iopart}

\pdfoutput=1
\usepackage{iopams}
\usepackage{graphicx}
\usepackage{epsfig}
\usepackage[latin1]{inputenc}
\usepackage{amssymb}

\begin{document}

\title{Polyakov Loops in Strongly-Coupled Plasmas with Gravity Duals}
\author{Jorge Noronha$^1$
\footnote{E-mail: noronha@phys.columbia.edu},
}

\address{$^1$Department of Physics, Columbia University, 538 West 120$^{th}$ Street, New York,
NY 10027, USA\\[0.2ex]
}

\begin{abstract}
We study the properties of the Polyakov loop in strongly-coupled gauge plasmas that are conjectured to be dual to five dimensional theories of gravity coupled to a nontrivial single scalar field. We find a gravity dual that can describe the thermodynamic properties and also the expectation value of the Polyakov loop in the deconfined phase of quenched $SU(3)$ QCD up to $3T_c$. 

\end{abstract}
%

\section{Introduction}\label{intro}

The expectation value of the Polyakov loop is an order parameter for the deconfinement phase transition in $SU(N_c)$ gauge theories without fermions \cite{polyakov} since $Z(N_c)$ symmetry is broken in the deconfined phase. At large $N_c$ the phase transition becomes a strong first-order phase transition \cite{largeNclattice} and the Polyakov loop jumps from zero to a finite value at $T_c$ (which is also known as the Gross-Witten point \cite{adrianrob}). We define the path-ordered Polyakov loop as
\begin{equation}
\bold{L} (\vec{x})=\frac{1}{N_c}{\rm Tr} \,P \,e^{i\int_{0}^{1/T} \hat{A}_{0}(\vec{x},\tau)d\tau}
\label{wilsonloop}
\end{equation}
where $\hat{A}^{\mu}$ is the non-Abelian gauge field operator and the trace is over the fundamental representation of $SU(N_c)$. According to the equation above, the loop can be seen as the trace of the propagator of an infinitely heavy quark in imaginary time. A less abstract way to understand this operator involves the connection between the loop and the free energy of a single, isolated (and infinitely massive) heavy quark in the medium \cite{McLerran:1980pk} $F_Q(T)$, i.e.,
\begin{equation}
\ell \equiv |\langle \bold{L} \rangle| = e^{-F_{Q}(T)/T}\,.
\label{definefreenergy}
\end{equation} 
Thus, in the confined phase of $SU(N_c)$ Yang-Mills (YM) theories the free energy diverges (because of Wilson's area law) and $\ell$ then vanishes. If dynamical fermions are added then $\ell$ is small but nonzero in the hadronic phase because $Q\bar{Q}$ pairs can be created thermally. The loop in QCD was computed in perturbative theory a long time ago and the leading correction to its asymptotic value (which is unity because of asymptotic freedom) is {\it positive} and of order $g^3$ \cite{Gava:1981qd}. This implies that in perturbation theory $\ell$ approaches unity from above. We will show in this paper that gauge theories with gravity duals that can mimic the current lattice predictions for the thermodynamics of pure glue can also describe $\ell$.

Because of cluster decomposition, $\ell$ can also be obtained from the disconnected piece of the Polyakov loop correlator 
\begin{equation}
\ \langle \bold{L}(r)\bold{L}(0)^{\dagger} \rangle = e^{-F_{Q\bar{Q}}(r,T)/T}\,,
\label{potential}
\end{equation} 
which is assumed to give the free energy, $F_{Q\bar{Q}}(r,T)$, associated with a heavy $Q\bar{Q}$ pair separated by a distance $r$ in the medium (for a recent review of quarkonia at finite temperature see \cite{Bazavov:2009us}).  

\section{Gravity Duals and the Breaking of Conformal Invariance}

The gauge/string duality states that the complicated non-Abelian dynamics of four dimensional gauge theories can be understood in terms of the physics of higher dimensional theories of gravity \cite{Maldacena:1997re,adsprescription}. This duality establishes that when the gauge theory is strongly-coupled the corresponding gravitational description is weakly-coupled, which makes this whole idea hard to falsify. However, there is in fact enough evidence that the main ideas behind this conjecture are right \cite{Aharony:1999ti}. 

The most well studied example of this correspondence is the duality between 4-dimensional $\mathcal{N}=4$ Supersymmetric Yang-Mills (SYM) and type IIB string theory on $AdS_5\otimes S_5$ \cite{Maldacena:1997re}. In this case, the gauge theory at large $N_c$ is in the regime of strong t'Hooft coupling $\lambda \gg 1$ and the typical ``stringy" coupling in the gravitational theory is $\alpha'/R^2=1/\sqrt{\lambda} \ll 1$, where $\sqrt{\alpha'}$ is the string length and $R$ is the radius of $AdS_5$. Here we refer the 5-dimensional spacetime in the gravity duals as the ``bulk" and the 4-dimensional flat spacetime where the gauge theory is defined as the ``boundary". Note that the boundary of $AdS_5$ is a flat 4d spacetime. Also, because in YM the only degrees of freedom are adjoint gluons, no other adjoint scalar fields in the gauge theory are included here and our working assumption is that the gravity dual theory may be properly defined in a 5 dimensions at least when it comes to the observables discussed in this paper.  

Gauge theories at finite temperature $T$ in general correspond to dual geometries with a black brane and Hawking temperature $T_{HW}=T$ \cite{Witten:1998zw}. $\mathcal{N}=4$ SYM is exactly conformal invariant in the vacuum and at finite temperature the ratio $p/T^4$ (where $p$ is the pressure) does not depend on $T$. Here we consider the work of Gubser and Nellore in \cite{Gubser:2008ny} that describes gravity duals where conformal invariance is broken in the infrared but restored when $T \to \infty$. This is described using a scalar field, $\phi$, that has a nontrivial dependence on the fifth dimension of the bulk spacetime. The gravitational action is
\begin{equation}
\ S=\frac{1}{2 k_5^2} \int d^5x\sqrt{-G}\left[\mathcal{R}-\frac{(\partial \phi)^2}{2}-V(\phi)\right]
\label{dilatonaction}
\end{equation}  
where the gravitational constant $k_5^2$ is a free parameter $\sim \mathcal{O}(1/N_c^2)$ and $V(\phi)$ defines the nontrivial profile displayed by $\phi$ in the bulk. Conformal invariance in the UV can be obtained when $\phi\to 0$ at the boundary ($u\to 0$) and
\begin{equation}  
\lim_{\phi\to 0} V(\phi)=-\frac{12}{R^2}+\frac{1}{2 R^2}\Delta(\Delta-4)\phi^2+\mathcal{O}(\phi^4),
\label{conditionboudnary}
\end{equation}
where the mass squared of the scalar $m_{\phi}^2 R^2=\Delta(\Delta-4)\geq -4$ according to the Breitenlohner-Freedman (BF) bound \cite{BFbound}. Here $\Delta$ is the UV scaling dimension of the operator that couples with $\phi$ at the boundary. Eq.\ (\ref{conditionboudnary}) implies that the spacetime will be asymptotically AdS$_5$ with radius $R$. 
        
The most general ansatz for the metric is \cite{Gubser:2008ny}
\begin{equation}
\ ds^2 = e^{2A(u)}(-f(u) dt^2+d\vec{x}^2)+e^{2B(u)}\frac{du^2}{f(u)}\,,\qquad \phi=\phi(u).
\label{metric}
\end{equation} 
($e^B$ has dimensions of length). The boundary of the bulk spacetime is located at $u=0$. Finite temperature effects are included by considering solutions of the classical equations of motion from Eq.\ (\ref{dilatonaction}) with a horizon at $u=u_h$ where $f(u_h)=0$. The functions above are assumed to be finite at $u_h$. We use a gauge where $\phi=u$ \cite{Gubser:2008ny}. Using this ansatz for the metric the entropy density is given by the area of the black brane horizon
\begin{equation}
s=2\pi\,\frac{e^{3A(u_h)}}{k_5^2} 
\label{bekenstein}
\end{equation}
while the Hawking temperature is (we use a prime for $d/du$)
\begin{equation}
T=\frac{e^{A(u_h)-B(u_h)}|f'(u_h)|}{4\pi} \,.
\label{hawking}
\end{equation} 
The class of $V(\phi)$ potentials considered here is of the form \cite{Gubser:2008ny} 
\begin{equation}
\ V(\phi)=\frac{-12\,\cosh \gamma \phi+b_1\phi^2}{R^2},
\label{potential}
\end{equation} 
where $\gamma$ is related to the speed of sound in the IR and $b_1$ determines $\Delta$ near the UV fixed point. The choice $\gamma=\sqrt{1/2}$ and $b_1=1.95$ gives $\Delta=3.3784$ and leads to a black hole solution with the thermodynamic properties shown in Fig.\ 1, which matches the lattice data for $c_s^2$ in $SU(3)$ YM above $T_c$, as computed by Boyd et al. \cite{Boyd:1996bx}. Note that even though the gravity calculations performed here require that $N_c\to \infty$ to avoid loop corrections, a comparison of our results to lattice data is still meaningful because, apart from a trivial rescaling, the thermodynamic properties of pure glue at $N_c=3$ and $N_c>3$ are very similar, as shown in \cite{largeNclattice}. Our calculations are done by computing numerically the entropy density using Eq.\ (\ref{bekenstein}). The other thermodynamic quantities are obtained from the entropy density via the thermodynamic identities and they also match the available data really well above $T_c$. 

\begin{figure}[h]
\centering
\includegraphics[width=7cm]{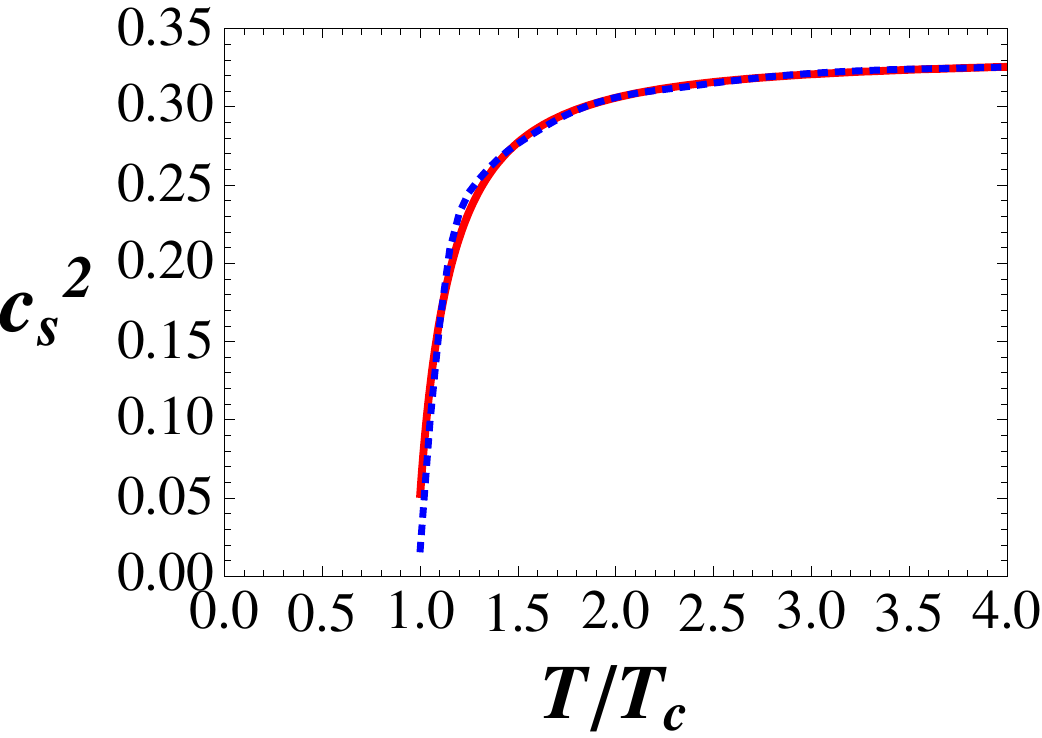} 
\caption{$c_s^2(T)$ in the gravity model (solid red line) compared to (an interpolation of) the lattice data (dotted line) from \cite{Boyd:1996bx}. }
\label{fig:cs2}
\end{figure}

\section{Polyakov Loops and the Gauge/String Duality}

When $N_c \to \infty$ and the radius of the background spacetime is much larger than the string length, an infinitely massive excitation in the fundamental representation of $SU(N_c)$ in the gauge theory at finite temperature is dual to a classical string in the bulk that goes from a probe D-brane at the boundary and to the black brane horizon \cite{Maldacena:1998im,Rey:1998bq,Bak:2007fk}. The endpoints of the strings are non-dynamical fundamental probes and they are equivalent to the infinitely heavy quark limit in quenched lattice gauge theory. 

The correlator of Polyakov loops in this approach is given by $\langle \ell (\vec{x})\ell^{*}(0)\rangle = Z_{string}$, where $Z_{string}$ is the full string generating functional that includes a sum over all the string worldsheets $\mathcal{D}$ whose boundaries describe a $Q\bar{Q}$ pair separated by a distance $r$ in the gauge theory \cite{Bak:2007fk}. Within this approximation, the string dynamics (in Euclidean space) is given by the classical Nambu-Goto (NG) action 
\begin{equation}
\ S_{NG}(\mathcal{D})=T_s\int_{\mathcal{D}} d^2\sigma \sqrt{{\rm det} \,h_s^{ab}}
\label{nambugotoaction}
\end{equation}
where $T_s=\frac{1}{2\pi \alpha'}$ is the string tension, $h_s^{ab}=G_s^{\mu\nu}\partial^{a}X^{\mu}\partial^{b}X^{\nu}$
($a,b=1,2$), $X^{\mu}=X^{\mu}(\tau,\sigma)$ is the embedding of the string in
the 5-dimensional spacetime, and $G_s^{\mu\nu}$ is the background bulk metric in the string frame. We assume here that the metric in these frames are related as follows: $G_s^{\mu\nu}=e^{\sqrt{2/3} \phi}G^{\mu\nu}$ \cite{gursoy,comment1}.    

The disconnected contribution to the correlator in theories such as (\ref{dilatonaction}) can be computed as follows. Using the gauge choice $X^{\mu}=(t,x(u),0,0,u)$ where the string endpoints at the boundary are at $x=r/2$ and $x=-r/2$, the regularized free energy of a single (static) heavy quark (up to a $T$ independent constant) above $T_c$ is
\begin{equation}
\ F_{Q}(T) = T_s \int_{0}^{u_h}du\left(\sqrt{M(u)}-\sqrt{M_0(u)}\right)-T_s \,u_h^{\frac{1}{\Delta-4}}
\label{singlefenergy}
\end{equation}
where $M(u)\equiv |G_{s\,00}|G_{s,\,uu}$ is a positive-defined function of $u$ that is regular at the horizon. Close to the boundary this function is $M_0(u)=\lim_{u\to 0}M(u)=u^{\frac{10-2\Delta}{\Delta-4}}/(4-\Delta)^2$ as long as $V$ obeys Eq.\ (\ref{conditionboudnary}). 
   
Instead of using Eq.\ (\ref{singlefenergy}), it is easier \cite{Noronha:2009ud} to work with its derivative $dF_{Q}/dT=\left(dF_{Q}/du_h\right) \,du_h/dT$. In fact, 
\begin{equation}
\frac{dF_{Q}}{du_h}= T_s \sqrt{M(u_h)} = T_s \,e^{\sqrt{\frac{2}{3}}\,u_h}\,e^{A(u_h)+B(u_h)} 
\label{dFduh}
\end{equation}
and one obtains that
\begin{equation}
\frac{dF_{Q}}{dT}=T_s\, e^{\sqrt{\frac{2}{3}}u_h}\, e^{A(u_h)+B(u_h)}\frac{du_h}{T d\ln T}\,.
\label{dFdT}
\end{equation}
Since the speed of sound squared in the plasma is $c_s^2=d \ln T/d \ln s $ we see that $du_h/d\ln s= c_s^2 du_h/d\ln T$. Thus, one can rewrite the equation above explicitly in terms of $c_s^2$
\begin{equation}
\frac{dF_{Q}}{dT}=T_s \,e^{\sqrt{\frac{2}{3}}u_h} \,e^{A(u_h)+B(u_h)}\frac{s}{T\,c_s^2}\frac{du_h}{ds}\,
\label{dFdT1}
\end{equation}
It is possible to show \cite{Noronha:2009ud} that in gravity duals governed by Eq.\ (\ref{dilatonaction}) the following equation holds
\begin{equation}
\frac{dF_{Q}}{dT}= 4\pi T_s \frac{e^{\sqrt{\frac{2}{3}}u_h(T)}}{V(u_h(T))}\frac{1}{c_s^2(T)} \,.
\label{dFdT1}
\end{equation}
This shows that $dF_{Q}/dT < 0$ (as expected from thermodynamics). Since $V(u_h)$ is well defined at finite $u_h$, one can see that $dF_{Q}/dT$ only diverges when $c_s \to 0$. Moreover, for confining gravity duals the magnitude of the jump in $\ell$ at $T_c$ is determined by how quickly $c_s\to 0$ at $T\to T_c^{+}$. In the case of a first-order transition it can be shown explicitly that the thermal width of heavy quarkonia at strong coupling derived in \cite{Noronha:2009da} vanishes below $T_c$. Continuous transitions where $\ell$ smoothly interpolates between zero and its high-T limit, such as the one that occur in QCD with dynamical fermions \cite{Bazavov:2009zn}, can also be obtained depending on the choice for $V(\phi)$. Because of our choice for $V$, the medium is always in the deconfined phase and $\ell$ can be evaluated at any $T$ using (\ref{dFdT1}) and there is no jump at $T_c$. 

The renormalized Polyakov loop in quenched $SU(3)$ QCD was computed some years ago on the lattice in \cite{Kaczmarek:2002mc} and more recently in \cite{Gupta:2007ax}. We calculated $\ell$ by integrating Eq.\ (\ref{dFdT1}) and, because of the regularization, $F_Q$ is determined only up to a constant whose value is chosen via a match to the lattice data at $T=1.03\, T_c$ (the first data point above $T_c$ in Ref.\ \cite{Gupta:2007ax} for $N_t=4$) for a given choice of $\alpha'$. Our results for the renormalized Polyakov loop are shown in Fig.\ 2 (with $\alpha'/R^2=3.5$) and one can see they agree very well with the lattice calculations from \cite{Gupta:2007ax} between $T/T_c=1-3$. At higher temperatures the curve starts to considerably deviates from the data. This may be due to the fact that the gauge theory considered here is not asymptotically free.  

\begin{figure}[h]
\centering
\includegraphics[width=7cm]{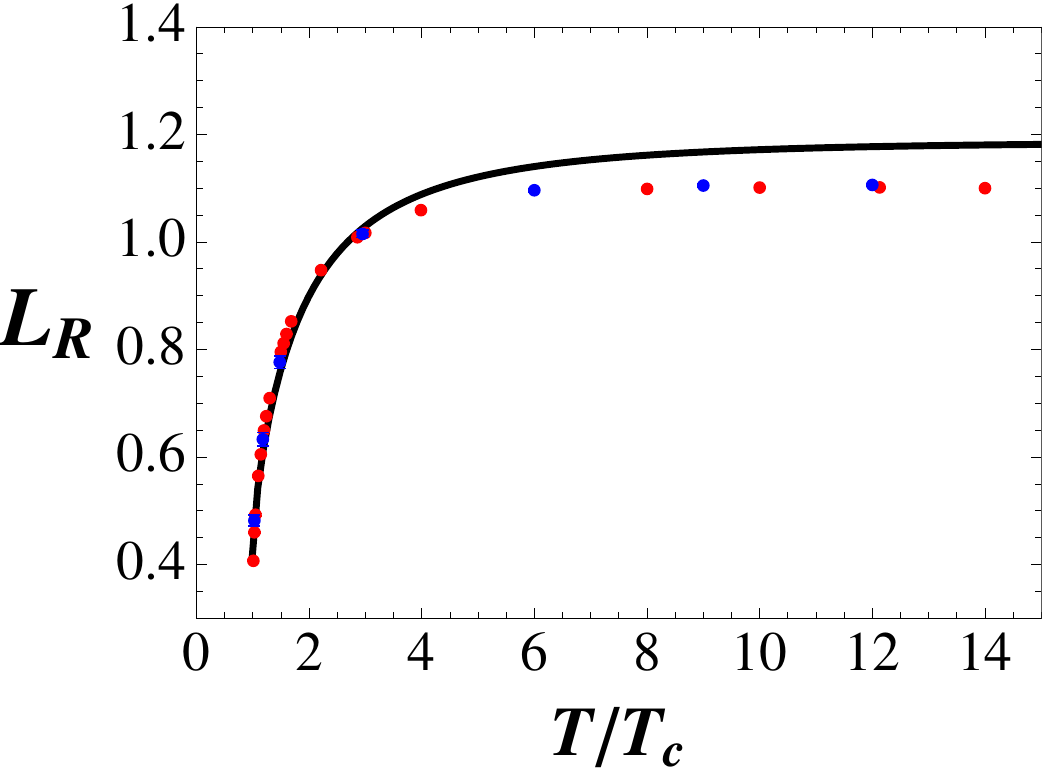} 
\caption{The renormalized Polyakov loop (solid black line) for the gravity dual in Eqs.\ (\ref{dilatonaction}) and (\ref{potential}) and the lattice data for quenched QCD with $N_t=4$ (red) and $N_t=8$ (blue) from Ref.\ \cite{Gupta:2007ax}.}
\label{fig:HS}
\end{figure}

\section{Conclusions}

We have shown that it is possible to describe the current lattice data for the thermodynamics and also the expectation value of the Polyakov loop of quenched SU(3) QCD using gravity duals where conformal invariance is softly broken by the presence of a nontrivial scalar field. The match gets worse as the temperature is increased, which may be a consequence of the fact that the theories considered here flow to a nontrivial fixed point (conformal invariance is restored when $T\to \infty$ but the system is still strongly interacting). It would be interesting to study examples where the gauge theory is confining \cite{gursoy} and also investigate the properties of the Polyakov loop and its correlator in this case. 

Note that in order to match the lattice data of the loop $\alpha'/R^2>1$, which indicates that higher-order derivative corrections \cite{higherderivative} to the gravity dual are probably going to be important in this type of analysis. These corrections were used in \cite{Noronha:2009ia} to study the dependence of $\ell$ with the shear viscosity to entropy density ratio, $\eta/s$, in superconformal strongly-coupled gauge theories. It would be interesting to extend this analysis to non-conformal gauge theories as well. In this case, $\eta$ and $\ell$ will be temperature dependent and one could obtain how $\eta$ varies with $\ell$ at strong t'Hooft coupling (this was computed at weak coupling in \cite{Hidaka:2008dr}).            

\section{Acknowledgements}

I would like to thank the organizers of SQM09 for a very enjoyable meeting and A.~Dumitru and R.~Pisarski for insightful discussions. This work was supported by the US-DOE Nuclear Science Grant No.\ DE-FG02-93ER40764. 

\section*{References}



\begin{thebibliography}{99}
\bibitem{comment1} Gursoy and collaborators \cite{gursoy} considered the more ``standard" notation where the string frame action is $S=\frac{1}{2 k_5^2} \int d^5x\sqrt{-G_s}e^{-2\Phi}\left[\mathcal{R}_s+4(\partial \Phi)^2+\ldots\right]$, which shows that this action was computed at tree level perturbation theory and, as such, there is a prefactor of $1/g_s^2$, where $g_s=e^\Phi$ is the string coupling constant. Thus, in this case $G_s^{\mu\nu}=e^{4\Phi/3} G^{\mu\nu}$. In our notation, $\Phi=\sqrt{3/8}\,\phi$.
\end{thebibliography}
\end{document}